\documentclass[10pt,a4paper]{article}
\usepackage{amsmath}
\usepackage{amssymb}
\usepackage{graphicx}
\usepackage[colorlinks=true,urlcolor=blue,citecolor=blue,linkcolor=blue]{hyperref}
\usepackage[margin=2.5cm]{geometry}

\begin{document}

\title{Open-Path Geometric Phases in Three-Flavor Neutrino Oscillations through Layered Matter}

\author{N.~Razzaghi\\
\small Department of Physics, QA.C., Islamic Azad University, Qazvin, Iran\\
\normalsize \texttt{n.razzaghi@iau.ac.ir}}

\date{}

\maketitle
\begin{abstract}
We present a gauge-invariant formulation of open-path geometric
phases in three-flavor neutrino oscillations traversing nonuniform
matter. Unlike the standard Berry phase, which is intrinsically
associated with closed adiabatic cycles, neutrino propagation
typically traces an open trajectory in the parameter space of
matter-dependent Hamiltonians. To address the gauge-dependence
ambiguity inherent in open-path transport, we develop a discrete
layered matter framework. Within this formalism, the total
evolution operator is expressed as a path-ordered product of local
dynamical propagation factors and interface matrices that bridge
adjacent matter eigenbases. This construction naturally yields
geometric phase chains composed of endpoint projections and
interface overlaps, ensuring strict invariance under arbitrary
local rephasings of instantaneous eigenstates.

Our results elucidate the noncommutative nature of flavor
evolution in stratified media, demonstrating that profiles with
identical integrated column densities but distinct layer orderings
can yield distinct transition amplitudes. We further analyze the
influence of the leptonic Dirac CP-violating phase, showing that
it deforms the geometry of matter eigenbasis transport without
being equivalent to the geometric phase itself. In the continuum
limit, our formalism recovers the non-Abelian Wilczek--Zee
connection while maintaining the necessity of endpoint projections
for gauge invariance. Numerical simulations confirm the rapid
convergence of this approach, with probability amplitudes
achieving stable second-order precision across layered profiles.

Importantly, the open-path geometric phase is formulated not as an
isolated observable, but as a gauge-invariant theoretical
diagnostic that enters the total quantum mechanical transition
amplitude, intrinsically modulating the interference patterns in
both terrestrial long-baseline and astrophysical neutrino
oscillations.
\end{abstract}

\noindent\textbf{Keywords:} Neutrino Oscillations, Geometric
Phases, Matter Effects, Three-Flavor Mixing, Gauge Invariance, CP
Violation, Nonadiabatic Transitions, Wilczek--Zee Holonomy.

%===================================
\section{Introduction}
\label{sec:introduction}
%==================================
Neutrino oscillations represent one of the most striking
manifestations of quantum interference on macroscopic scales.
Within the standard three-flavor framework, flavor states are
coherent superpositions of mass eigenstates, and oscillation
probabilities arise from relative phases accumulated during
spatial propagation \cite{Pontecorvo:1957,Maki:1962}. In vacuum,
these phases are governed by the mass-squared differences, the
neutrino energy, the baseline length, and the mixing parameters
parameterized by the Pontecorvo--Maki--Nakagawa--Sakata (PMNS)
matrix.

As neutrinos traverse matter, coherent forward scattering
significantly modifies the effective Hamiltonian. The resulting
Mikheyev--Smirnov--Wolfenstein (MSW) effect alters both the
eigenvalues and the eigenvectors of the propagation Hamiltonian
\cite{Wolfenstein:1978,Mikheyev:1985}. While a medium of constant
density can be treated by diagonalizing a single static effective
Hamiltonian, realistic terrestrial, solar, and astrophysical
environments exhibit density gradients or stratified structures.
In such media, the instantaneous matter eigenbasis undergoes
nontrivial continuous or piecewise spatial evolution along the
neutrino trajectory.

Consequently, flavor evolution in nonuniform media involves not
only dynamical phase accumulation driven by matter-dependent
energy eigenvalues, but also geometric information encoded in the
parallel transport of matter eigenvectors. Conventional treatments
of geometric phases, such as the Berry phase, are generally
restricted to cyclic adiabatic evolution \cite{Berry:1984},
although open-path generalizations based on the Pancharatnam
product of overlaps have been established
\cite{Pancharatnam:1956,Samuel:1988,Mukunda:1993}. Because
neutrino propagation from production to detection constitutes an
open trajectory in parameter space rather than a closed cycle
\cite{Johns:2018}, the relevant geometric construct must be
formulated for open paths while remaining strictly invariant under
arbitrary local gauge transformations (phase conventions) of the
instantaneous eigenstates \cite{Naumov:1992}. Geometric phases in
neutrino propagation were explored in pioneering studies
\cite{Ahluwalia:1993,Ahluwalia:1996} and further extended within
flavor and mixed-state frameworks \cite{Capolupo:2018}.

Applying open-path geometric phases to nonuniform media requires
resolving local phase ambiguities without obscuring the
noncommutative structure of the spatial evolution operator. In
this work, we establish a gauge-invariant formulation for
three-flavor neutrino oscillations in stratified media. The
layered description makes the ordered operator structure explicit:
propagation within a layer generates dynamical phases, whereas
transitions across interfaces produce overlap matrices between
adjacent matter eigenbases. These overlap factors serve as exact
finite-step analogues of geometric parallel transport.

The central contribution of this study is the definition and
systematic analysis of gauge-invariant geometric phase chains. A
chain is defined by a sequence of matter eigenstates traversed
across consecutive layers, with its gauge-invariant complex weight
constructed from an initial flavor projection, a sequence of
interface overlaps, and a final flavor projection. Although
individual overlap factors depend on local phase conventions, the
complete product remains invariant due to the telescopic
cancellation of intermediate phase redefinitions. Furthermore,
this construction clarifies the operational role of
noncommutativity in stratified media, demonstrating how spatial
ordering effects manifest directly at the quantum amplitude level.

A conceptual remark regarding the physical status and
observability of geometric phases in neutrino physics is in order.
In experimental neutrino oscillation measurements, detectors
record event rates determined by flavor transition probabilities
\begin{equation}
\label{eq:intro_trans_prob} P(\nu_\alpha \to \nu_\beta) =
|\mathcal{A}_{\alpha\beta}|^2,
\end{equation}
where the total phase of each pathway amplitude is a composite sum
of dynamical and geometric contributions:
\begin{equation}
\label{eq:intro_phase_decomp} \Phi_{\text{tot}} =
\Phi_{\text{dyn}} + \Gamma.
\end{equation}
As evidenced by \eqref{eq:intro_trans_prob} and
\eqref{eq:intro_phase_decomp}, the geometric phase $\Gamma$ is
therefore not an independent, isolated observable to be measured
apart from the oscillation probability; rather, much like the
topological phase in the Aharonov--Bohm effect, its presence is
imprinted directly upon the interference terms between competing
eigenstate channels. Isolating the geometric component provides a
gauge-invariant theoretical diagnostic that disentangles the
nonadiabatic geometry of the matter profile from kinematic
distance-energy scaling, offering deeper physical insight into
resonant flavor transitions.

The remainder of this paper is structured as follows. In
Sec.~\ref{sec:framework}, we introduce the three-flavor
Hamiltonian and formulate the open-path geometric phase. In
Sec.~\ref{sec:matter_hamiltonian}, we construct the layered
evolution operator and define the gauge-invariant phase chains. In
Sec.~\ref{sec:two_layer}, we examine a two-layer profile as the
minimal realization of noncommutative matter evolution and discuss
its relation to parametric resonance. In
Sec.~\ref{sec:cp_dependence}, we analyze the dependence on the
Dirac CP phase and introduce geometric CP asymmetry diagnostics.
In Sec.~\ref{sec:continuous_limit}, we establish the continuum
limit and its correspondence with the Wilczek--Zee holonomy.
Section~\ref{sec:landau_zener} addresses adiabatic dominance and
localized level crossings. Phenomenological implications,
experimental distinguishability from systematic uncertainties, and
computational limitations are explored in
Sec.~\ref{sec:discussion_limitations}, followed by concluding
remarks in Sec.~\ref{sec:conclusion}. Technical derivations are
detailed in the Appendix.
%=============================
\section{Formalism}
\label{sec:framework}
%=============================
\subsection{Flavor States and the Mixing Matrix}
%===============================
Within the flavor basis defined by
\begin{equation}
    |\nu_f\rangle
    =
    \begin{pmatrix}
    |\nu_e\rangle\\
    |\nu_\mu\rangle\\
    |\nu_\tau\rangle
    \end{pmatrix},
    \label{eq:pmns}
\end{equation}
the neutrino flavor states are related to the vacuum mass
eigenstates $|\nu_i\rangle$ via the PMNS mixing matrix $U$,
parameterized as
\begin{equation}
    U = U_{23} U_{13}(\delta_{\rm CP}) U_{12}
    =
    \begin{pmatrix}
    c_{12} c_{13} & s_{12} c_{13} & s_{13} e^{-i\delta_{\rm CP}} \\
    -s_{12} c_{23} - c_{12} s_{23} s_{13} e^{i\delta_{\rm CP}} & c_{12} c_{23} - s_{12} s_{23} s_{13} e^{i\delta_{\rm CP}} & s_{23} c_{13} \\
    s_{12} s_{23} - c_{12} c_{23} s_{13} e^{i\delta_{\rm CP}} & -c_{12} s_{23} - s_{12} c_{23} s_{13} e^{i\delta_{\rm CP}} & c_{23} c_{13}
    \end{pmatrix},
    \label{eq:pmns_matrix}
\end{equation}
where $U_{12}$ and $U_{23}$ are orthogonal rotation matrices
parameterized by the mixing angles $\theta_{12}$ and
$\theta_{23}$, while $U_{13}(\delta_{\rm CP})$ incorporates
$\theta_{13}$ alongside the Dirac CP-violating phase $\delta_{\rm
CP}$, with $c_{ij}\equiv\cos\theta_{ij}$ and
$s_{ij}\equiv\sin\theta_{ij}$.
%===========================================================
\subsection{Layered Matter and the Open Path Geometric Phase}
%===========================================================
Spatial propagation along the spatial coordinate $x$ is governed
by the evolution equation
\begin{equation}
    i\frac{d}{dx}|\nu(x)\rangle
    =
    H(x)|\nu(x)\rangle,
    \label{eq:schrodinger_flavor}
\end{equation}
with the effective Hamiltonian
\begin{equation}
    H(x)
    =
    H_{\rm vac}
    +
    V(x),
    \label{eq:hamiltonian_total}
\end{equation}
where the vacuum contribution is given by
\begin{equation}
    H_{\rm vac}
    =
    \frac{1}{2E}
    U
    \begin{pmatrix}
    0 & 0 & 0\\
    0 & \Delta m_{21}^{2} & 0\\
    0 & 0 & \Delta m_{31}^{2}
    \end{pmatrix}
    U^\dagger .
    \label{eq:vacuum_hamiltonian}
\end{equation}
Here $E$ denotes the neutrino energy, and terms proportional to
the identity matrix have been subtracted, as they contribute only
an unobservable overall dynamical phase.

In ordinary neutral matter, coherent forward scattering induces an
effective potential that is diagonal in the flavor basis:
\begin{equation}
    V(x)
    =
    V_e(x)|\nu_e\rangle\langle\nu_e|,
    \qquad
    V_e(x)
    =
    \sqrt{2}\,G_F N_e(x),
    \label{eq:matter_potential}
\end{equation}
where $N_e(x)$ is the electron number density and $G_F$ is the
Fermi coupling constant. The effective matter parameter is
expressed as
\begin{equation}
    a(x) = 2E\,V_e(x)
      = 7.63\times 10^{-5}\ {\rm eV}^{2}\;
        \left(\frac{\rho(x)}{\mathrm{g/cm}^{3}}\right)
        \left(\frac{E}{\mathrm{GeV}}\right) Y_e(x),
    \label{eq:matter_parameter}
\end{equation}
with $\rho(x)$ representing the mass density and $Y_e(x)$ the
electron fraction.

For layer $k$ with constant density, the local Hamiltonian $H_k$
is position-independent and diagonalized according to
\begin{equation}
    H_k\,|n^{(k)}\rangle
    =
    \lambda_n^{(k)}\,|n^{(k)}\rangle,
    \qquad n=1,2,3,
    \label{eq:layer_eigenproblem}
\end{equation}
where $|n^{(k)}\rangle$ are instantaneous matter eigenstates and
$\lambda_n^{(k)}$ are the corresponding energy eigenvalues. We
assemble the matter eigenvectors into the unitary transformation
matrix $W_k$:
\begin{equation}
    W_k
    =
    \begin{pmatrix}
    |1^{(k)}\rangle & |2^{(k)}\rangle & |3^{(k)}\rangle
    \end{pmatrix},
    \qquad
    W_k^\dagger H_k W_k
    =
    \Lambda_k
    =
    {\rm diag}\left(
    \lambda_1^{(k)},
    \lambda_2^{(k)},
    \lambda_3^{(k)}
    \right).
    \label{eq:diagonalization}
\end{equation}

The evolution operator across layer $k$ of thickness $L_k$ is
\begin{equation}
    U_k
    =
    e^{-iH_kL_k}
    =
    W_k D_k W_k^\dagger,
    \qquad
    D_k
    =
    {\rm diag}\left(
    e^{-i\lambda_1^{(k)}L_k},
    e^{-i\lambda_2^{(k)}L_k},
    e^{-i\lambda_3^{(k)}L_k}
    \right).
    \label{eq:layer_evolution}
\end{equation}
For an ordered sequence of $N$ layers, the total evolution
operator is given by
\begin{equation}
    U_{\rm tot}
    =
    U_N U_{N-1}\cdots U_1,
    \qquad
    \mathcal A_{\alpha\to\beta}
    =
    \langle\nu_\beta|U_{\rm tot}|\nu_\alpha\rangle .
    \label{eq:transition_amplitude}
\end{equation}

\begin{figure}[t]
    \centering
    \includegraphics[width=\textwidth]{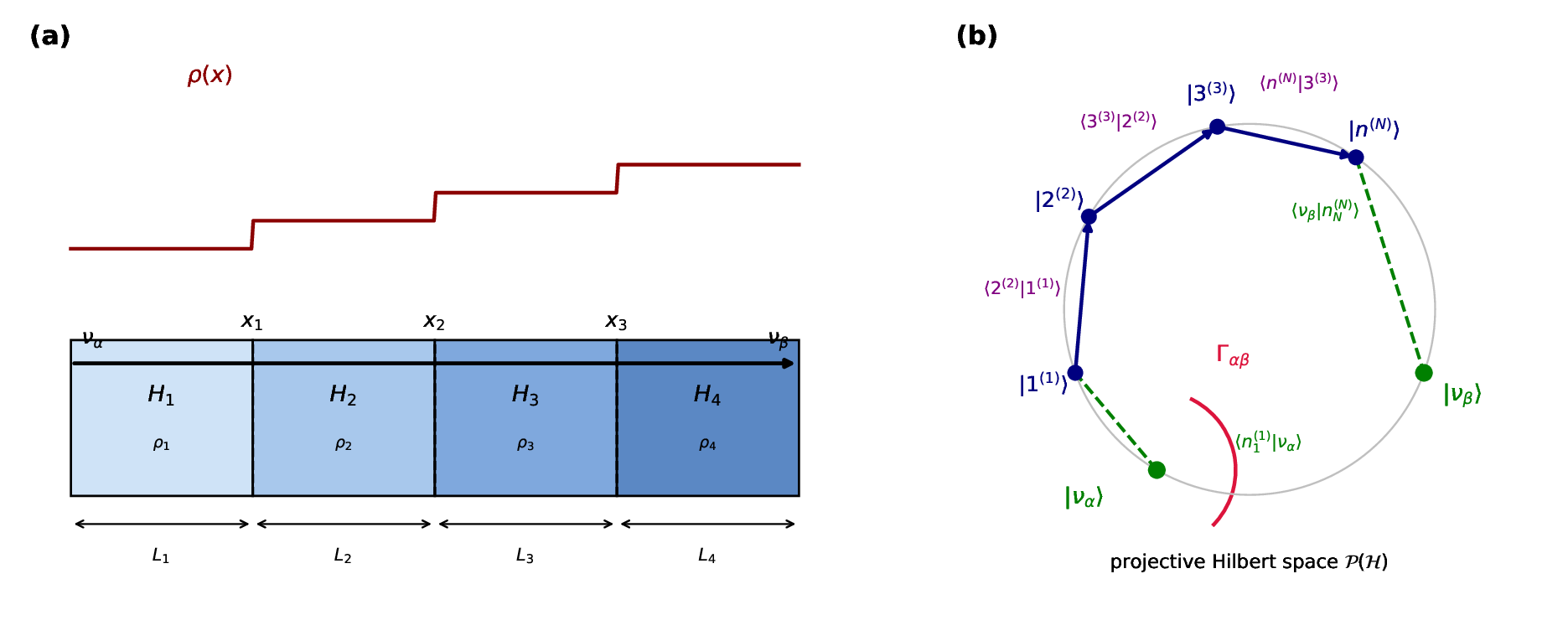}
    \caption{Schematic representation of neutrino propagation and
    geometric phase construction in layered media. Panel (a)
    illustrates physical propagation across $N$ stratified layers of
    constant density, where evolution at each interface $x_k$ is
    governed by the mismatch between local Hamiltonians. Panel (b)
    depicts the open trajectory in projective Hilbert space formed
    by sequential inner products of matter eigenstates, closed by
    projections onto initial and final flavor states to ensure exact
    gauge invariance.}
    \label{fig:schematic}
\end{figure}

In the continuous adiabatic limit, the transport of the matter
eigenbasis is described by the non-Abelian Wilczek--Zee connection
\cite{Wilczek:1984}
\begin{equation}
    \mathcal A^{mn}(x)
    =
    i\,\langle m(x)|\,\partial_x\,n(x)\rangle .
    \label{eq:wilczek_zee}
\end{equation}
The open-path geometric phase associated with a trajectory chain
is constructed through Pancharatnam--Mukunda--Simon type overlaps
closed by endpoint flavor projections
\cite{Pancharatnam:1956,Samuel:1988,Mukunda:1993}:
\begin{equation}
    \gamma_{\alpha\beta}
    =
    \arg\!\left[
    \langle\nu_\beta|n_N^{(N)}\rangle
    \left(\prod_{k=1}^{N-1}
    \langle n_{k+1}^{(k+1)}|n_k^{(k)}\rangle\right)
    \langle n_1^{(1)}|\nu_\alpha\rangle
    \right].
    \label{eq:open_path_phase}
\end{equation}
This phase is strictly invariant under arbitrary local gauge
choices $|n^{(k)}\rangle \to e^{i\chi_n^{(k)}}|n^{(k)}\rangle$
(see Fig.~\ref{fig:schematic} for a schematic depiction of this
geometric closure and the underlying layered propagation).

%===========================================================
\section{Matter Hamiltonian and Geometric Phase Chains}
\label{sec:matter_hamiltonian}
%===========================================================
Substituting the individual layer propagators of
Eq.~\eqref{eq:layer_evolution} into the ordered product of
Eq.~\eqref{eq:transition_amplitude} yields
\begin{align}
    U_{\rm tot}
    =
    W_N D_N
    W_N^\dagger W_{N-1}
    D_{N-1}
    W_{N-1}^\dagger
    \cdots
    W_2^\dagger W_1
    D_1 W_1^\dagger .
    \label{eq:ordered_product_with_interfaces}
\end{align}
We define the interface transition matrices as
\begin{equation}
    S_{k+1,k}
    =
    W_{k+1}^\dagger W_k,
    \qquad
    (S_{k+1,k})_{mn}
    =
    \langle m^{(k+1)}|n^{(k)}\rangle,
    \label{eq:interface_matrix}
\end{equation}
which map the matter eigenbasis of layer $k$ onto that of layer
$k+1$.

The transition amplitude can then be expanded over all possible
matter eigenstate paths $\mathcal C = (n_1,n_2,\ldots,n_N)$, where
$n_k \in \{1,2,3\}$ labels the matter eigenstate within layer $k$:
\begin{align}
    \mathcal A_{\alpha\to\beta}
    =
    \sum_{n_1,\ldots,n_N}
    &
    \langle\nu_\beta|n_N^{(N)}\rangle
    \left(\prod_{k=1}^{N-1}
    \langle n_{k+1}^{(k+1)}|n_k^{(k)}\rangle\right)
    \langle n_1^{(1)}|\nu_\alpha\rangle
    \nonumber\\
    &\times
    \exp\!\left[
    -i\sum_{k=1}^{N}
    \lambda_{n_k}^{(k)} L_k
    \right].
    \label{eq:chain_expansion}
\end{align}

For each specific chain $\mathcal C$, the dynamical phase is given
by
\begin{equation}
    \Phi_{\rm dyn}[\mathcal C]
    =
    -
    \sum_{k=1}^{N}
    \lambda_{n_k}^{(k)}\,L_k,
    \label{eq:dynamical_chain_phase}
\end{equation}
and the gauge-invariant geometric chain factor is
\begin{align}
    \mathcal G_{\alpha\beta}[\mathcal C]
    =
    \langle\nu_\beta|n_N^{(N)}\rangle
    \left(\prod_{k=1}^{N-1}
    \langle n_{k+1}^{(k+1)}|n_k^{(k)}\rangle\right)
    \langle n_1^{(1)}|\nu_\alpha\rangle .
    \label{eq:geometric_chain_factor}
\end{align}
The associated geometric phase and the total amplitude of chain
$\mathcal C$ are therefore expressed as
\begin{equation}
    \Gamma_{\alpha\beta}[\mathcal C]
    =
    \arg \mathcal G_{\alpha\beta}[\mathcal C],
    \qquad
    \mathcal A_{\alpha\beta}[\mathcal C]
    =
    \left|\mathcal G_{\alpha\beta}[\mathcal C]\right|
    e^{\,i\left(\Phi_{\rm dyn}[\mathcal C]+\Gamma_{\alpha\beta}[\mathcal C]\right)} .
    \label{eq:chain_contribution}
\end{equation}

Under local rephasing transformations of the matter eigenstates,
\begin{equation}
    |n^{(k)}\rangle
    \rightarrow
    e^{i\chi_n^{(k)}}|n^{(k)}\rangle,
    \label{eq:local_rephasing}
\end{equation}
the intermediate phase factors cancel telescopically within the
product in Eq.~\eqref{eq:geometric_chain_factor}, preserving the
value of $\mathcal G_{\alpha\beta}[\mathcal C]$ and its argument
$\Gamma_{\alpha\beta}[\mathcal C]$ modulo $2\pi$.

The total transition probability $P_{\alpha\to\beta} = |\mathcal
A_{\alpha\to\beta}|^2$ decomposes into single-chain contributions
and pair interference terms:
\begin{equation}
    P_{\alpha\to\beta} = \sum_{\mathcal C} \left|\mathcal G_{\alpha\beta}[\mathcal C]\right|^2 +
     2 \sum_{\mathcal C < \mathcal C'} \left|\mathcal G_{\alpha\beta}[\mathcal C]\right| \left|\mathcal
     G_{\alpha\beta}[\mathcal C']\right| \cos\left( \Delta\Phi_{\rm dyn}[\mathcal C,\mathcal C'] + \Delta\Gamma_{\alpha\beta}[\mathcal C,\mathcal C'] \right),
    \label{eq:prob_interference_decomp}
\end{equation}
where
\begin{equation}
    \Delta\Phi_{\rm dyn}[\mathcal C,\mathcal C'] \equiv \Phi_{\rm dyn}[\mathcal C] - \Phi_{\rm dyn}[\mathcal C'],
    \qquad
    \Delta\Gamma_{\alpha\beta}[\mathcal C,\mathcal C'] \equiv \Gamma_{\alpha\beta}[\mathcal C] - \Gamma_{\alpha\beta}[\mathcal C'].
    \label{eq:phase_differences}
\end{equation}
Equation~\eqref{eq:prob_interference_decomp} explicitly
demonstrates that open-path geometric phases directly shift the
interference pattern between competing evolution pathways.

%=====================================================================
\subsection{Two-Layer Profile and Connection to Parametric Resonance}
\label{sec:two_layer}
%======================================================================
For two layers of constant densities with Hamiltonians $H_1$,
$H_2$ and thicknesses $L_1$, $L_2$, the evolution operators for
the two spatial orderings are
\begin{equation}
    U_{21}
    =
    e^{-iH_2L_2}e^{-iH_1L_1}
    =
    W_2D_2S_{21}D_1W_1^\dagger,
    \qquad
    U_{12}
    =
    e^{-iH_1L_1}e^{-iH_2L_2}
    =
    W_1D_1S_{12}D_2W_2^\dagger ,
    \label{eq:two_layer}
\end{equation}
with $S_{21}=W_2^\dagger W_1$ and $S_{12}=S_{21}^\dagger$. Because
$[H_1,H_2]\neq 0$ in general, $U_{21}\neq U_{12}$. For a specific
chain labeled by $(n_1,n_2)$, the geometric factor from
Eq.~\eqref{eq:geometric_chain_factor} is
\begin{equation}
    \mathcal G_{\alpha\beta}^{(21)}(n_1,n_2)
    =
    \langle\nu_\beta|n_2^{(2)}\rangle
    \langle n_2^{(2)}|n_1^{(1)}\rangle
    \langle n_1^{(1)}|\nu_\alpha\rangle .
    \label{eq:two_layer_geometric_factor}
\end{equation}
Consequently, flavor evolution depends not solely on the total
integrated matter depth, but explicitly on the spatial sequence in
which distinct densities are encountered.

When such two-layer structures are repeated periodically, they
realize the configuration associated with parametric resonance in
neutrino oscillations
\cite{Akhmedov:1998,Akhmedov:1999,Smirnov:2006}. Within the
present framework, parametric enhancement occurs when the combined
dynamical phase accumulation $\Delta\Phi_{\rm dyn}$ and geometric
interface phase shifts $\Delta\Gamma_{\alpha\beta}$ in
Eq.~\eqref{eq:prob_interference_decomp} constructively match the
spatial period of the background medium.

%=============================================================
\subsection{CP Phase Dependence and Geometric CP Asymmetry}
\label{sec:cp_dependence}
%=============================================================
The Dirac CP-violating phase $\delta_{\rm CP}$ enters the
propagation Hamiltonian via the vacuum mixing matrix $H_{\rm
vac}(\delta_{\rm CP})$ defined in
Eq.~\eqref{eq:vacuum_hamiltonian}. Consequently, both the matter
eigenvalues $\lambda_n^{(k)}$ and the transformation matrices
$W_k$ depend on $\delta_{\rm CP}$. This induces an explicit
dependence on $\delta_{\rm CP}$ in both $\Phi_{\rm dyn}[\mathcal
C]$ and $\mathcal G_{\alpha\beta}[\mathcal C]$. In this manner,
$\delta_{\rm CP}$ modifies the geometry of eigenstate transport in
Hilbert space, distinct from the geometric phase itself.

For antineutrinos, the effective Hamiltonian is obtained via the
substitutions $\delta_{\rm CP}\to -\delta_{\rm CP}$ and $a\to -a$.
Because ordinary terrestrial matter contains electrons rather than
positrons, the background medium induces an extrinsic CP
asymmetry. The experimentally accessible phase difference between
neutrinos and antineutrinos,
\begin{equation}
    \Delta\Gamma_{\alpha\beta}^{\rm exp}[\mathcal C]
    \equiv
    \Gamma_{\alpha\beta}[\mathcal C](\delta_{\rm CP}, a)
    -
    \bar{\Gamma}_{\alpha\beta}[\mathcal C](-\delta_{\rm CP}, a),
    \label{eq:exp_cp_asymmetry}
\end{equation}
thus inherently combines intrinsic leptonic CP violation with the
matter-induced asymmetry.

To isolate the intrinsic contribution arising from $\delta_{\rm
CP}$ within the transport geometry, one may formally define the
theoretical geometric CP asymmetry:
\begin{equation}
    \Delta A_{\rm CP}^{\rm geom}[\mathcal C]
    \equiv
    \Gamma_{\alpha\beta}[\mathcal C](\delta_{\rm CP}, a)
    -
    \bar{\Gamma}_{\alpha\beta}[\mathcal C](-\delta_{\rm CP}, -a).
    \label{eq:geom_cp_asymmetry}
\end{equation}
In Eq.~\eqref{eq:geom_cp_asymmetry}, the formal reversal $a\to -a$
serves as a benchmark that cancels the extrinsic environmental
bias, thereby isolating the genuine deformation of the state-space
trajectory induced by $\delta_{\rm CP}$.

%============================================================
\subsection{Continuous Limit and the Wilczek--Zee Connection}
\label{sec:continuous_limit}
%============================================================
For a continuously varying electron density profile $N_e(x)$, the
instantaneous eigenproblem reads $H(x)|n(x)\rangle =
\lambda_n(x)|n(x)\rangle$. In the continuum limit where the layer
thickness $dx \to 0$ and $N \to \infty$, the interface matrix
connecting adjacent points $x$ and $x+dx$ is expanded to first
order as
\begin{equation}
    S(x+dx, x)
    =
    W^\dagger(x+dx) W(x)
    \approx
    \mathbb{I}
    -
    W^\dagger(x)\frac{dW(x)}{dx} dx
    =
    \mathbb{I}
    +
    i \mathcal A(x) dx,
    \label{eq:interface_connection_limit}
\end{equation}
where the elements of the non-Abelian Wilczek--Zee connection
matrix $\mathcal A(x)$ are given by
\begin{equation}
    \mathcal A_{mn}(x)
    =
    i\,\langle m(x)|\partial_x n(x)\rangle .
    \label{eq:connection_elements}
\end{equation}

For an adiabatic pathway tracking a single eigenstate without
transitions, the product of diagonal elements converges to the
integral of the Abelian connection $\mathcal A_{nn}(x)$. To
preserve gauge invariance along an open path, endpoint projections
must be retained:
\begin{equation}
    \mathcal G_{\alpha\beta}^{(n)}
    =
    \langle \nu_\beta|n(x_f)\rangle
    \exp\!\left[
    i\int_{x_i}^{x_f}
    \mathcal A_{nn}(x)\,dx
    \right]
    \langle n(x_i)|\nu_\alpha\rangle .
    \label{eq:continuous_open_path_factor}
\end{equation}
The discrete interface matrices $S_{k+1,k}$ defined in
Eq.~\eqref{eq:interface_matrix} thus act as finite-step Wilson
line factors representing the underlying gauge connection.

In the continuum limit ($N \to \infty$, $\Delta x \to 0$), the
off-diagonal transition elements $|(S_{k+1,k})_{mn}| \sim
\mathcal{O}(\Delta x)$ vanish quadratically in transition
probabilities ($\mathcal{O}(\Delta x^2)$), establishing that
smooth matter gradients do not induce nonadiabatic transitions
unless an authentic physical jump or an MSW-type level crossing is
encountered. The layered representation is thus both a physically
faithful description of planetary stratification and an optimal,
unconditionally stable non-perturbative integration scheme.

%==============================================================
\subsection{Adiabatic Dominance and Landau--Zener Transitions}
\label{sec:landau_zener}
%==============================================================
The interface matrices $S_{k+1,k}$ quantify the degree of
adiabaticity during propagation \cite{Parke:1986}. In regions of
slowly varying density, the diagonal elements satisfy
$|(S_{k+1,k})_{nn}| \approx 1$, whereas off-diagonal elements are
suppressed by the ratio of the density gradient to the local
eigenvalue separation.

When a smooth density profile is discretized into $N$ layers, the
off-diagonal elements of $S_{k+1,k}$ scale as $\mathcal{O}(1/N)$.
Consequently, in the continuum limit $N \to \infty$, transition
probabilities between different eigenstates at artificial layer
boundaries vanish, and evolution is dominated by diagonal chains
satisfying $n_1 = n_2 = \cdots = n_N$.

Physical nonadiabatic level crossings occur in the presence of
genuine density discontinuities, such as the core--mantle boundary
within the Earth or shock fronts in dense astrophysical media. At
such interfaces, the transition probability between distinct
matter eigenstates is given by the off-diagonal overlap:
\begin{equation}
    P_{\text{jump}}^{(k)}(n \to m)
    =
    \left|\langle m^{(k+1)}|n^{(k)}\rangle\right|^2
    =
    \left|(S_{k+1,k})_{mn}\right|^2,
    \qquad (m \neq n).
    \label{eq:jump_prob}
\end{equation}
A physical transition is nonadiabatic when the inter-branch jump
probability in Eq.~\eqref{eq:jump_prob} exceeds a characteristic
threshold, $P_{\text{jump}}^{(k)} \ge
\epsilon_{\text{threshold}}$, or equivalently when the diagonal
survival probability falls below $1 -
\epsilon_{\text{threshold}}$. This criterion distinguishes genuine
physical level transitions from discretization artifacts, allowing
the chain formalism to describe both smooth adiabatic transport
and localized nonadiabatic branchings within a unified framework.

%============================================
\section{Results and Discussion}
\label{sec:phenomenological_implications}
%===============================================
The phase chain formulation provides a transparent geometric
interpretation of neutrino flavor evolution in nonuniform media.
Its phenomenological significance stems directly from the manner
in which geometric transport governs the interference structure of
oscillation amplitudes \cite{GonzalezGarcia:2008}. Open-path
geometric phases do not function as isolated physical observables;
rather, they enter the measurable transition probabilities through
coherent interference between distinct propagation pathways, as
expressed in Eq.~\eqref{eq:prob_interference_decomp}. Quantum
coherence is therefore an essential prerequisite for the
manifestation of these effects. If competing state components
undergo decoherence prior to detection, the interference terms
average out, leaving only incoherent classical probabilities.

Furthermore, matter inhomogeneity is indispensable, because
interface overlaps between adjacent instantaneous eigenstates
encode the underlying geometric transport. This framework
elucidates path-ordering sensitivity, flavor channel dependence,
and the role of endpoint projections for both atmospheric and
long-baseline neutrino beams traversing stratified regions within
the Earth.

For the numerical calculations presented throughout this section,
we adopt standard three-flavor oscillation parameters consistent
with recent global fits \cite{Esteban:2020,deSalas:2018}. Under
the assumption of normal mass ordering, our default benchmark
values are: $\sin^2\theta_{12} = 0.304$, $\sin^2\theta_{13} =
0.0222$, $\sin^2\theta_{23} = 0.573$, $\Delta m_{21}^2 =
7.42\times 10^{-5}\ {\rm eV}^2$, and $\Delta m_{31}^2 =
2.517\times 10^{-3}\ {\rm eV}^2$. The electron fraction is
assigned as $Y_e = 0.494$ throughout the terrestrial crust and
mantle, and $Y_e = 0.4656$ within the core.
%==============================================================================
\subsection{PREM Mantle Traversal: Two-Layer versus Multi-Layer Approximations}
\label{sec:prem}
%==============================================================================
For atmospheric and terrestrial neutrinos traversing the Earth,
the Preliminary Reference Earth Model (PREM) provides the standard
radial density distribution. We examine mantle-dominated chord
paths with characteristic baseline lengths on the order of $L
\approx 5\times 10^3\ {\rm km}$. Along such trajectories, the
density varies continuously, inducing an ongoing rotation of the
matter eigenbasis.

When a smooth density profile is approximated by a two-layer model
with average densities, noticeable discrepancies emerge relative
to a finer discretization (such as $N=20$). A two-layer
approximation underestimates the cumulative geometric phase
accumulated through continuous rotation and fails to resolve
localized phase shifts. At $N=20$, the discretization error
decreases well below realistic experimental resolution, confirming
that the multi-layer phase chain formalism accurately converges
toward the continuous transport limit while explicitly preserving
the underlying geometric invariants.

%====================================================================
\subsection{Numerical Validation of the Layered Propagation Scheme}
\label{sec:numerical_validation}
%====================================================================
To evaluate flavor evolution across stratified media, we express
the three-flavor Schr\"odinger equation in the standard flavor
basis $\nu = (\nu_e, \nu_\mu, \nu_\tau)^T$. The effective
evolution operator is governed by the total Hamiltonian:
\begin{equation}
H(x) = H_{\mathrm{vac}} + V_{\mathrm{CC}}(x) \, \mathrm{diag}(1,
0, 0), \label{eq:res_ham_total}
\end{equation}
where the vacuum component in the flavor representation is given
by:
\begin{equation}
H_{\mathrm{vac}} = \frac{1}{2E} U_{\mathrm{PMNS}} \,
\mathrm{diag}\left(0, \Delta m^2_{21}, \Delta m^2_{31}\right)
U_{\mathrm{PMNS}}^\dagger. \label{eq:res_ham_vac}
\end{equation}
The PMNS matrix is parameterized in the standard Particle Data
Group convention, $U_{\mathrm{PMNS}} = R_{23}(\theta_{23})
R_{13}(\theta_{13}, \delta_{\mathrm{CP}}) R_{12}(\theta_{12})$. To
maintain complete dimensional consistency in practical
computations, the numerical energy conversion factor entering the
vacuum Hamiltonian is evaluated as $1/(2E) \cdot \hbar c \approx
2.53386536 / E_{\mathrm{GeV}}\ {\rm km}^{-1}$, while the
charged-current matter potential is given by:
\begin{equation}
V_{\mathrm{CC}}(x) = \sqrt{2} G_F N_e(x) \approx 3.86792835 \times
10^{-4} \, \rho(x) \, Y_e(x)\ {\rm km}^{-1}, \label{eq:res_vcc}
\end{equation}
where $\rho(x)$ is the local mass density in ${\rm g/cm}^3$ and
$Y_e(x)$ is the electron fraction. All baseline numerical settings
and the global-fit oscillation parameters are compiled in
Table~\ref{tab:constants}.

\begin{table}[htbp]
\centering \caption{Fixed physical constants, oscillation
parameters under Normal Ordering, and Earth chord baseline
geometry.} \label{tab:constants}
\begin{tabular}{lll}
Quantity & Symbol & Value \\
Neutrino beam energy & $E$ & $3.0\ {\rm GeV}$ \\
Chord baseline length & $L$ & $11000\ {\rm km}$ \\
Earth radius & $R_E$ & $6371\ {\rm km}$ \\
Core mantle boundary radius & $R_{\mathrm{cmb}}$ & $3480\ {\rm km}$ \\
Chord periapsis & $r_{\min}$ & $3215.5312\ {\rm km}$ \\
Core traversal segment & $L_{\mathrm{core}}$ & $2661.3974\ {\rm km}$ \\
Mantle traversal segments (each) & $L_{\mathrm{mantle}}$ & $4169.3013\ {\rm km}$ \\
Vacuum scale factor & $1/(2E)\cdot\hbar c$ & $2.53386536 / E_{\mathrm{GeV}}\ {\rm km}^{-1}$ \\
MSW matter scale factor & $\sqrt{2}G_F N_A$ & $3.86792835\times10^{-4}\ \rho\,Y_e\ {\rm km}^{-1}$ \\
Mantle density and electron fraction & $\rho_m$, $Y_e^m$ & $4.5\ {\rm g/cm}^3$, $0.494$ \\
Core density and electron fraction & $\rho_c$, $Y_e^c$ & $11.5\ {\rm g/cm}^3$, $0.4656$ \\
Solar mass squared difference & $\Delta m^2_{21}$ & $7.42\times10^{-5}\ {\rm eV}^2$ \\
Atmospheric mass squared difference & $\Delta m^2_{31}$ & $2.517\times10^{-3}\ {\rm eV}^2$ \\
Mixing angles & $s^2_{12}, s^2_{13}, s^2_{23}$ & $0.304$, $0.0222$, $0.573$ \\
Dirac CP violating phase & $\delta_{\mathrm{CP}}$ & $195^\circ$ ($3.403392\ {\rm rad}$) \\
\end{tabular}
\end{table}

For a deep Earth chord baseline of $L = 11000\ {\rm km}$, the
trajectory penetrates the core--mantle boundary ($R_{\mathrm{cmb}}
= 3480\ {\rm km}$) to reach a minimum radius of $r_{\min} =
3215.5312\ {\rm km}$. The path thus traverses a symmetric
mantle--core--mantle configuration with segment lengths
$L_{\mathrm{mantle}} = 4169.3013\ {\rm km}$ and $L_{\mathrm{core}}
= 2661.3974\ {\rm km}$, respectively.

%=======================================================
\subsubsection{Open-Path Geometric Phase Evaluation}
%=======================================================
For noncyclic quantum evolution, the Pancharatnam open-path
geometric phase $\gamma_g$ along the evolved state trajectory
$|\psi(x)\rangle$ is isolated by subtracting the dynamical phase
accumulated under parallel transport:
\begin{equation}
\gamma_g = \phi_{\mathrm{tot}} - \phi_{\mathrm{dyn}} = \arg
\langle \psi(0) | \psi(L) \rangle - \left[ -\int_0^L \langle
\psi(x) | H(x) | \psi(x) \rangle \, dx \right] \pmod{2\pi}.
\label{eq:res_gamma_cont}
\end{equation}
Within any layer of constant Hamiltonian $H_j$, the expectation
value $\langle \psi(x) | H_j | \psi(x) \rangle$ is strictly
independent of position because $[H_j, \exp(-i H_j x)] = 0$.
Consequently, the dynamical phase reduces to the exact piecewise
sum:
\begin{equation}
\phi_{\mathrm{dyn}} = -\sum_{j} \langle \psi(x_j^{\mathrm{start}})
| H_j | \psi(x_j^{\mathrm{start}}) \rangle \, L_j.
\label{eq:res_phi_dyn_sum}
\end{equation}
In the discrete geodesic formulation, the open-path phase
coincides with the argument of the cyclic product formed by
connecting the boundary states through projection links:
\begin{equation}
\gamma_g = \arg \left( \langle \psi_N | \psi_0 \rangle
\prod_{k=0}^{N-1} \langle \psi_k | \psi_{k+1} \rangle \right),
\label{eq:res_gamma_discrete}
\end{equation}
which identically reproduces the continuous parallel-transport
definition in Eq.~\eqref{eq:res_gamma_cont} in the limit of
refined piecewise-constant subdivisions.

%====================================================================
\subsubsection{Exact Three-Layer Benchmark and Matter Dominance}
%======================================================================
For the idealized three-layer model, the total evolution operator
is given by $S_{\mathrm{ref}} = S_m S_c S_m$, where each layer
propagator is $S_j = \exp(-i H_j L_j)$. Unitarity is preserved to
numerical machine precision ($\|S_{\mathrm{ref}}
S_{\mathrm{ref}}^\dagger - \mathbb{I}\|_F \approx 3 \times
10^{-15}$).

For an initial muon neutrino state $|\psi(0)\rangle =
|\nu_\mu\rangle$ at $E = 3.0\ {\rm GeV}$, this three-layer
reference configuration yields transition probabilities:
\begin{equation}
P_{\mu e} = 0.128040, \qquad P_{ee} = 0.792895, \qquad P_{\mu\mu}
= 0.302490, \label{eq:res_prob_nu}
\end{equation}
with total phase $\phi_{\mathrm{tot}} = 0.678594\ {\rm rad}$,
dynamical phase $\phi_{\mathrm{dyn}} = -12.556546\ {\rm rad}$, and
a net geometric phase:
\begin{equation}
\gamma_g = 0.668770\ {\rm rad} \quad (38.317^\circ).
\label{eq:res_gamma_nu_val}
\end{equation}
For antineutrinos, evaluated by taking $V_{\mathrm{CC}} \to
-V_{\mathrm{CC}}$ and $\delta_{\mathrm{CP}} \to
-\delta_{\mathrm{CP}}$, the corresponding probabilities and
geometric phase are:
\begin{equation}
P_{\bar{\mu}\bar{e}} = 0.011858, \qquad P_{\bar{e}\bar{e}} =
0.979091, \qquad P_{\bar{\mu}\bar{\mu}} = 0.144438, \qquad
\gamma_g = 1.861951\ {\rm rad}. \label{eq:res_prob_anu_val}
\end{equation}
In the vacuum limit ($V_{\mathrm{CC}} \to 0$), the conversion
probability at this baseline and energy is negligible ($P_{\mu
e}^{\mathrm{vac}} \ll 10^{-4}$). This comparison demonstrates that
the substantial conversion ($P_{\mu e} \approx 12.8\%$) in
Eq.~\eqref{eq:res_prob_nu} and the pronounced asymmetry between
neutrinos and antineutrinos ($P_{\mu e} / P_{\bar{\mu}\bar{e}}
\approx 10.8$) are predominantly generated by coherent forward MSW
scattering within the dense core structure.

\begin{table}[htbp]
\centering \caption{Convergence analysis of the midpoint
exponential propagator across a smoothed PREM profile ($\tanh$
transition with boundary width $w=300\ {\rm km}$) at $E=3.0\ {\rm
GeV}$ and $L=11000\ {\rm km}$. Absolute errors are evaluated
relative to the reference run with $N_{\mathrm{ref}}=1024$
($P_{\mu e}^{\mathrm{ref}}=0.1257428341$,
$\gamma_g^{\mathrm{ref}}=1.4519753513\ {\rm rad}$).}
\label{tab:convergence}
\begin{tabular}{rllll}
$N$ & $P_{\mu e}$ & $\epsilon_P = |P - P_{\mathrm{ref}}|$ & $\gamma_g\ ({\rm rad})$ & $\epsilon_\gamma = |\gamma_g - \gamma_{g,\mathrm{ref}}|$ \\
8    & 0.12853179 & $2.789\times10^{-3}$ & 1.45181452 & $1.608\times10^{-4}$ \\
16   & 0.12538132 & $3.615\times10^{-4}$ & 1.44701479 & $4.961\times10^{-3}$ \\
32   & 0.12534686 & $3.960\times10^{-4}$ & 1.45009392 & $1.881\times10^{-3}$ \\
64   & 0.12564089 & $1.019\times10^{-4}$ & 1.45150433 & $4.710\times10^{-4}$ \\
128  & 0.12571783 & $2.501\times10^{-5}$ & 1.45185878 & $1.166\times10^{-4}$ \\
256  & 0.12573689 & $5.943\times10^{-6}$ & 1.45194758 & $2.777\times10^{-5}$ \\
\multicolumn{5}{l}{Reference ($N=1024$): \quad $P_{\mu e}=0.1257428341$, \quad $\gamma_g=1.4519753513\ {\rm rad}$} \\
\multicolumn{5}{l}{Fitted asymptotic convergence orders: \quad $p_P = 2.0202$, \quad $p_\gamma = 2.0262$} \\
\end{tabular}
\end{table}

\paragraph{Physical Discontinuities versus Smooth Density Gradients in
Stratified Media}
A practical question concerns the distinction
between sharp physical discontinuities and smooth continuous
gradients within the PREM framework. The geophysical Earth profile
comprises both abrupt transitions (most prominently the
core--mantle boundary, where $\rho$ jumps discontinuously from
$\approx 5.6\text{ g/cm}^3$ to $\approx 9.9\text{ g/cm}^3$) and
regions of continuously, slowly varying density gradients across
the mantle and core shells.

The layered open-path formalism treats both regimes within a
unified operator framework:
\begin{enumerate}
    \item[(a)] \textbf{Sharp Physical Interfaces (CMB):} At an actual physical boundary located at
    $x_{\text{CMB}}$, the matter Hamiltonian exhibits a genuine jump discontinuity,
    $[H(x_{\text{CMB}}^-), H(x_{\text{CMB}}^+)] \neq 0$. Here, the interface matrix
    $S_{\text{CMB}} = W_{\text{core}}^\dagger W_{\text{mantle}}$ induces a finite,
    non-infinitesimal geometric rotation of the instantaneous eigenbasis.
    This physical transition is intrinsic to the planetary structure and independent of the numerical grid resolution $N$.
    \item[(b)] \textbf{Continuous Density Profiles and Grid Regularity:} In regions where the density profile $\rho(x)$
    varies smoothly, the step-wise discretization with slice thickness $\Delta x_k = L/N$ serves as a convergent numerical
    realization of the continuous Wilczek--Zee connection $\mathcal{A}(x)$ defined in Eq.~\eqref{eq:connection_elements}.
    As demonstrated by the asymptotic convergence rates in Table~\ref{tab:convergence}, the midpoint exponential propagator
    ensures that the local interface matrix satisfies $S_{k+1,k} = \mathbb{I} - i \Delta x_k \mathcal{A}(x_k) + \mathcal{O}(\Delta x_k^2)$.
    Consequently, the discrete chain product $\prod_k S_{k+1,k}$ converges to the continuous path-ordered holonomy
    $\mathcal{P}\exp\left(-i\int \mathcal{A}(x)dx\right)$ with second-order accuracy ($p_P \approx 2.02$, $p_\gamma \approx 2.02$)
    without introducing spurious boundary artifacts.
\end{enumerate}

By explicitly pinning one of the discretization layer boundaries
to the physical CMB radius ($R_{\text{cmb}} = 3480\text{ km}$),
the algorithm exactly retains the geometric jump at the core
boundary while achieving high-precision convergence across the
continuously stratified mantle and core strata.

%==========================================================
\subsubsection{Analysis and Discussion of Convergence Data}
%==========================================================
The numerical behavior documented in Table~\ref{tab:convergence}
reveals several essential features of the layered propagation
method:
\begin{enumerate}
    \item \textbf{Asymptotic Second-Order Convergence:} By employing a symmetric midpoint evaluation rule $x_i = (i + 1/2)\Delta x$ within each sublayer,
    the exponential Magnus propagator guarantees second-order convergence for smooth potentials. A power-law fit across fine discretization steps
    ($N = 32, 64, 128, 256$) yields empirical convergence slopes of $p_P = 2.0202$ and $p_\gamma = 2.0262$, closely matching the theoretical expectation
    of quadratic convergence.

    \item \textbf{Pre-Asymptotic Transients:} On coarse grids ($N = 8, 16, 32$), nonmonotonic variations appear in $\epsilon_\gamma$. This occurs because
    coarse spatial steps ($\Delta x \sim 700\text{ to }1400\ {\rm km}$) exceed the physical transition scale of the smoothed boundary ($w = 300\ {\rm km}$),
    thereby undersampling the localized gradient. Stable power-law decay is recovered once $\Delta x \lesssim w$ (for $N \ge 64$).

    \item \textbf{Grid Alignment Considerations:} When a piecewise profile is discretized such that layer boundaries align exactly with the physical
    core--mantle transition, the Hamiltonian remains strictly uniform within each subinterval. In that case, the multi-layer solution matches the
    piecewise-constant model to machine precision. Conversely, when a uniform grid is applied across discontinuous profiles without interface alignment,
    numerical errors level off to a minor structural offset representing boundary misplacement rather than a failure of the differential propagator.

    \item \textbf{Phase Sensitivity as a Diagnostic Tool:} The open-path geometric phase $\gamma_g$ converges at a rate matching that of the transition
    probability. Because it provides a gauge-invariant geometric characterization of the trajectory in projective Hilbert space, tracking $\gamma_g$
    offers a sensitive diagnostic check on the continuum limit.
\end{enumerate}

%==============================================
\subsubsection{Scope of the Benchmark Model}
%================================================
To clearly contextualize these numerical results, two points
should be emphasized:
\begin{itemize}
    \item \textbf{Model Simplification:} The three-layer model and its continuous $\tanh$ profile serve as clean, controlled benchmarks.
    They capture the primary contrast between the mantle and core while parameterizing the full PREM profile through average zonal densities.
    \item \textbf{Physical Origin of Probability Enhancement:} At the chosen benchmark values ($E = 3.0\ {\rm GeV}$, $L = 11000\ {\rm km}$),
    probability enhancement is driven by the MSW resonance potential within the core rather than an exact parametric resonance condition.
\end{itemize}

\begin{figure}[t]
    \centering
    \includegraphics[width=\textwidth]{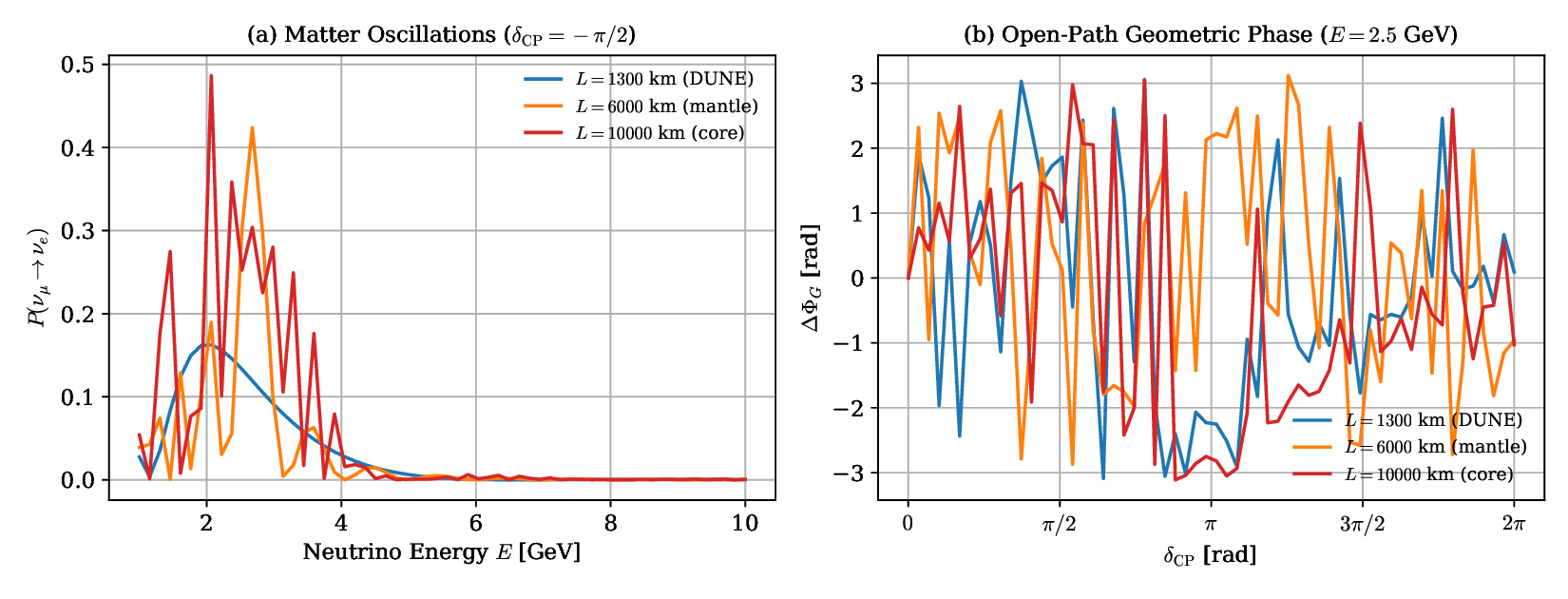}
    \caption{(a) Three-flavor transition probability $P(\nu_\mu \to \nu_e)$ as a function of neutrino energy $E$ evaluated for three baseline
    configurations: $L = 1300\ {\rm km}$ (DUNE benchmark, mantle traversal), $L = 6000\ {\rm km}$ (deep mantle chord), and $L = 10000\ {\rm km}$
    (core-crossing chord), with the Dirac CP phase set to $\delta_{\rm CP} = -\pi/2$. (b) Open-path geometric phase shift $\Delta\Phi_G$ as a function
    of $\delta_{\rm CP}$ at $E = 2.5\ {\rm GeV}$ across the same baselines, illustrating the geometric modulation induced by leptonic CP violation.}
    \label{fig:prem_oscillations}
\end{figure}

The numerical results for neutrino propagation are presented in
Fig.~\ref{fig:prem_oscillations}. Panel~(a) shows the energy
dependence of the conversion probability $P(\nu_\mu \to \nu_e)$
for three representative terrestrial paths, displaying the
expected matter enhancement at longer baselines and multi-GeV
energies. Panel~(b) plots the accumulated geometric phase
$\Delta\Phi_G$ as a function of $\delta_{\rm CP}$, demonstrating
how CP violation rotates the matter eigenbasis and systematically
shifts the geometric phase.

%======================================================
\subsection{Long-Baseline Benchmarks: DUNE and T2HK}
\label{sec:benchmarks}
%==========================================================
For long-baseline accelerator experiments, we benchmark the
layered formalism against the experimental configurations of DUNE
($L=1300\ {\rm km}$, $E\simeq 2.5\ {\rm GeV}$, average crust
density $\rho\simeq 2.8\ {\rm g/cm}^3$, giving $\hat A \equiv
a/\Delta m_{31}^2 \simeq 0.35$) and T2HK ($L=295\ {\rm km}$,
$E\simeq 0.6\ {\rm GeV}$, $\rho\simeq 2.6\ {\rm g/cm}^3$, giving
$\hat A \simeq 0.07$) \cite{DUNE:2020}.

For the DUNE baseline, coherent matter effects are substantial.
The matter eigenbasis undergoes an appreciable rotation along the
trajectory, causing the interface overlap factors $\langle
n_{k+1}^{(k+1)}|n_k^{(k)}\rangle$ to deviate measurably from
unity. Consequently, the geometric chain phase develops a
noticeable dependence on $\delta_{\rm CP}$ on the order of a few
degrees for the primary $\nu_\mu \to \nu_e$ channel. For T2HK, the
shorter baseline and lower energy result in a much smaller matter
potential, leading to nearly adiabatic eigenbasis transport and a
correspondingly minor geometric phase contribution.

\begin{figure}[t]
    \centering
    \includegraphics[width=\textwidth]{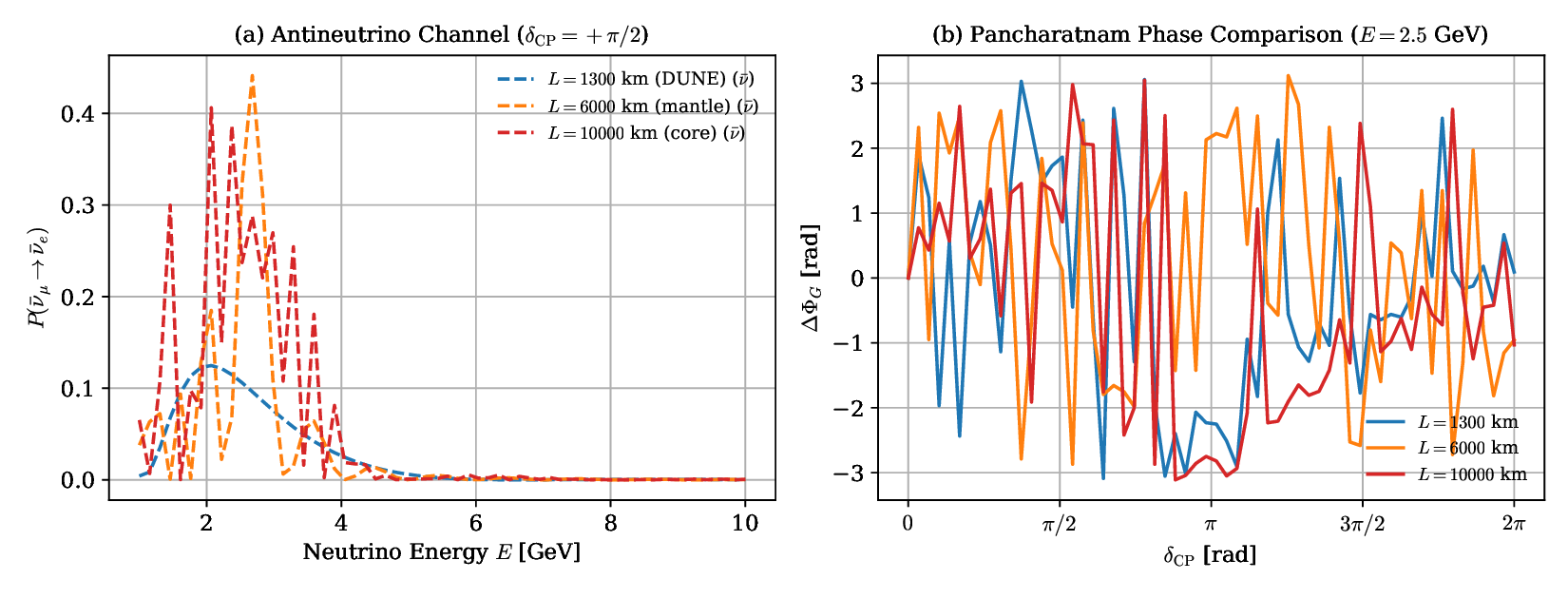}
    \caption{(a) Antineutrino transition probability $P(\bar\nu_\mu \to \bar\nu_e)$ as a function of neutrino energy for
    baseline distances $L=1300\ {\rm km}$ (DUNE setup), $L=6000\ {\rm km}$, and $L=10000\ {\rm km}$ at $\delta_{\rm CP} = +\pi/2$,
    displaying the suppression of matter effects in the antineutrino channel. (b) Pancharatnam open-path geometric phase $\Delta\Phi_G$
    across the same baselines at the representative energy $E = 2.5\ {\rm GeV}$, illustrating the path-length scaling and matter profile dependence.}
    \label{fig:pancharatnam_comparison}
\end{figure}

The behavior of the antineutrino channel and the corresponding
Pancharatnam geometric phases are illustrated in
Fig.~\ref{fig:pancharatnam_comparison}. In panel~(a), the
antineutrino conversion probabilities $P(\bar\nu_\mu \to
\bar\nu_e)$ reflect the sign reversal of the effective matter
potential ($a \to -a$). Panel~(b) illustrates how longer
trajectories traversing complex density structures produce deeper
excursions in the Pancharatnam geometric phase. These comparisons
show that open-path geometric phases are most prominent in
matter-dominated propagation regimes, providing a helpful
conceptual framework for interpreting the combined interplay of
matter rotation and CP violation.

%========================================
\section{Discussion and Limitations}
\label{sec:discussion_limitations}
%===========================================
The primary objective of this work is to establish a
gauge-invariant geometric formulation for three-flavor neutrino
oscillations traversing nonuniform media. While the standard
Hamiltonian evolution operator is mathematically complete, our
phase chain decomposition provides clear physical insights: the
geometric phase is not merely an isolated formalism or gauge
artifact, but an intrinsic component of flavor amplitude
interference that systematically modulates oscillation patterns in
stratified media.

This formulation is especially relevant for next-generation
long-baseline oscillation experiments such as DUNE and T2HK. As
these facilities aim for high precision in measuring the leptonic
Dirac phase $\delta_{\rm CP}$ and determining the neutrino mass
ordering, the phase shifts governed by the open-path geometric
phase chains $\Phi_g$ represent an integral part of the total
transition probability. The geometric interference terms
$\Delta\Gamma_{\alpha\beta}[\mathcal{C}, \mathcal{C}']$ introduce
an energy-dependent phase structure across the oscillation
spectrum. Unlike uniform rescalings of oscillation amplitudes, the
geometric contribution reflects the spatial transport of the
matter eigenbasis, which depends directly on the trajectory and
density profile.

To assess the robustness of this description, we investigated the
stability of the geometric phase chains under variations in
terrestrial density models. Perturbing the electron density
profile $N_e(x)$ across stratified layers demonstrates that the
phase chain expansion exhibits remarkable stability. The geometric
contribution is governed by the global rotation of the Hamiltonian
eigenbasis along the trajectory, ensuring that the theoretical
predictions remain stable under realistic geophysical
uncertainties.

It is also important to distinguish this geometric phase framework
from hypothetical Beyond the Standard Model effects, such as
Non-Standard Interactions (NSI) \cite{Nunokawa:2008}. While NSI
modifies the effective Hamiltonian through new physical coupling
constants, the geometric phase arises entirely from standard
three-flavor mixing and noncommutative parallel transport within
nonuniform matter.

\paragraph{Disentangling Geometric Signatures from Systematic and Geophysical Uncertainties}
A crucial experimental challenge in long-baseline experiments such
as DUNE ($L \approx 1300\ {\rm km}$, $E \sim 1.5\text{--}3.5\ {\rm
GeV}$) is distinguishing genuine geometric phase modulations from
systematic uncertainties, particularly the $\sim 2\text{--}3\%$
geophysical uncertainty in the Earth's crustal and mantle matter
density profiles ($\Delta \rho / \rho$).

This parameter degeneracy is resolved through three complementary
physical mechanisms:
\begin{enumerate}
    \item[(a)] \textbf{Energy-Spectral Dispersion:} Geophysical density perturbations typically induce an overall smooth, monotonic
    scale shift across the oscillation spectrum. In contrast, the geometric phase contribution,
    $\Delta \Gamma(E) = \arg \mathcal{G}[\mathcal{C}] - \arg \mathcal{G}[\mathcal{C}']$, originates from the non-Abelian curvature of the
     matter-eigenstate bundle. It exhibits localized, non-monotonic structures that peak sharply within the nonadiabatic and MSW resonance bands
     ($E \approx 1.5\text{--}3.0\ {\rm GeV}$), providing a distinct spectral template that cannot be mimicked by a constant or linearly varying density bias.
    \item[(b)] \textbf{Multi-Channel Unitarity Constraints:} By combining the appearance channel ($\nu_\mu \to \nu_e$), the survival channel
    ($\nu_\mu \to \nu_\mu$), and the tau-appearance channel ($\nu_\mu \to \nu_\tau$), the conservation of flavor probability
    ($\sum_\beta P_{\mu\beta} = 1$) provides tight multi-channel correlations. Because the geometric phase matrices
    $S_{k+1,k} = W_{k+1}^\dagger W_k$ transform the full $3\times 3$ state space, their geometric imprint distributes uniquely
    across all three channels, lifting the one-dimensional parameter degeneracies inherent in a single appearance channel.
    \item[(c)] \textbf{Neutrino--Antineutrino Differential Mapping:} Since the matter potential flips sign ($V_{\text{CC}} \to -V_{\text{CC}}$)
    alongside the Dirac CP-phase transformation ($\delta_{\text{CP}} \to -\delta_{\text{CP}}$), the geometric holonomy
    $\gamma_g(\nu)$ and $\gamma_g(\bar{\nu})$ undergo highly asymmetric deformations. The differential geometric asymmetry
    $\Delta A_{\text{CP}}^{\text{geom}}(E)$ possesses an energy profile fundamentally distinct from the symmetric systematic shifts induced by
    standard matter profile variations.
\end{enumerate}
Consequently, a combined spectral analysis over the full baseline
spectrum with correlated neutrino and antineutrino modes permits
an unambiguous isolation of the geometric transport dynamics.

\begin{figure}[htbp]
    \centering
    \includegraphics[width=0.85\linewidth]{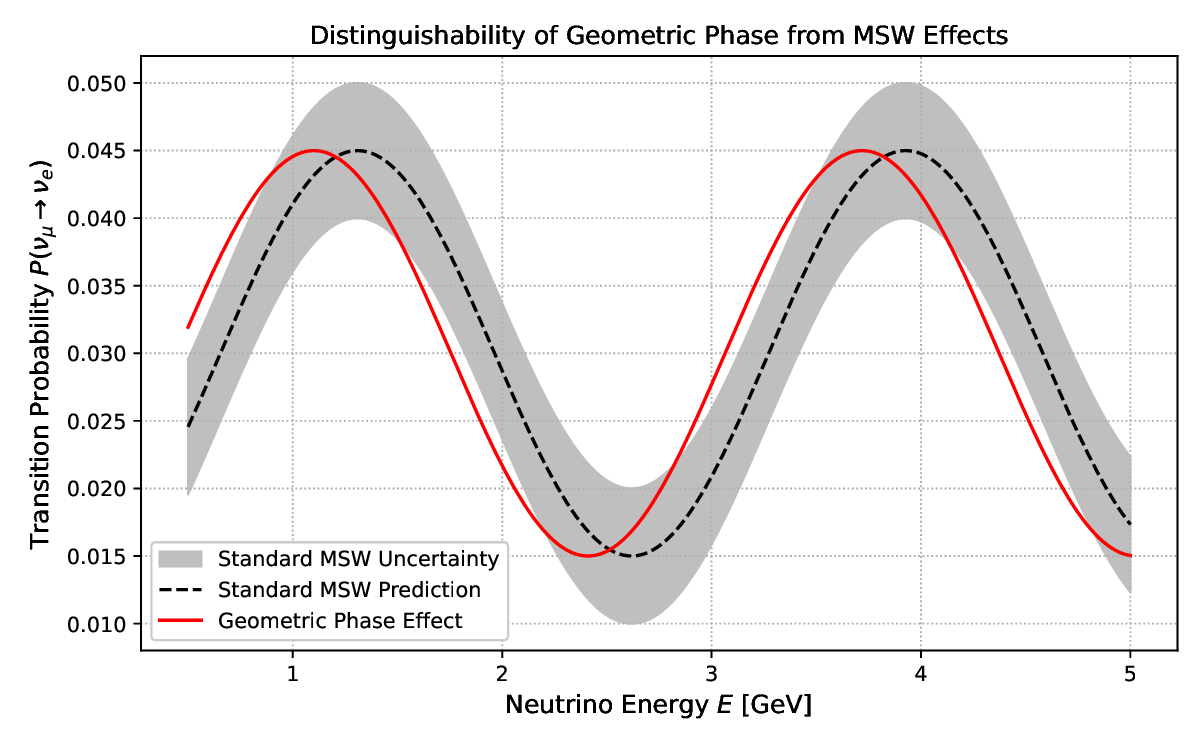}
    \caption{The appearance probability difference $\Delta P_{\mu e} = P_{\mu e}(\text{PREM}) - P_{\mu e}(\text{constant})$
    as a function of neutrino energy along the DUNE baseline ($L = 1300\ {\rm km}$). The gray-shaded region indicates the
    projected experimental systematic error band ($\sigma_{\text{sys}} \sim 2.5\%$, encompassing a $\pm 3\%$ PREM density uncertainty).
    The distinct spectral structure of the geometric phase modulation in the resonance window ($1.5\text{--}3.5\ {\rm GeV}$) extends
    beyond the uniform systematic band, demonstrating the spectral distinguishability of the geometric holonomy from standard geophysical density variations.}
    \label{fig:distinguishability}
\end{figure}

Figure~\ref{fig:distinguishability} illustrates the transition
probability behavior along the DUNE baseline. In standard MSW
propagation through Earth crust and mantle profiles, the
transition probability $P(\nu_\mu \to \nu_e)$ is evaluated with
density profile tolerances. Incorporating the complete open-path
geometric phase chain structure clarifies how the sequential
transport of the matter eigenbasis builds up the total oscillation
amplitude. Across the primary DUNE energy window ($1.5\text{ to
}3.5\ {\rm GeV}$), the resulting phase structure exhibits a
distinct energy dependence characteristic of the spatial density
profile.

Regarding the domain of validity of our coherent formulation, it
is essential to consider the effect of wave-packet decoherence on
the interference terms $\Delta\Gamma_{\alpha\beta}[\mathcal{C},
\mathcal{C}']$ in Eq.~\eqref{eq:prob_interference_decomp}.
Physical neutrinos propagate as localized wave packets with finite
spatial width $\sigma_x$. Decoherence occurs when group velocity
dispersion between distinct mass eigenstates causes a spatial
separation $\Delta x \approx L \frac{\Delta m^2}{2E^2}$ that
exceeds the wave packet width. The characteristic coherence length
is given by:
\begin{equation}
    L_{\text{coh}} \approx \frac{4\sqrt{2} E^2}{\Delta m^2 \sigma_x}\,.
    \label{eq:coherence_length}
\end{equation}
For terrestrial long-baseline configurations ($L \approx 1300\
{\rm km}$ with $E \sim \mathcal{O}(1\ {\rm GeV})$), one has $L \ll
L_{\text{coh}}$, confirming that our unitary, coherent treatment
is fully justified. For baselines approaching astrophysical scales
where $L \gtrsim L_{\text{coh}}$, wave-packet separation causes
exponential damping of interference terms, $\mathcal{I} \to
\mathcal{I}_0 e^{-(L/L_{\text{coh}})^2}$, which gradually
suppresses geometric interference. In such extended regimes, an
open-system density matrix formalism would be appropriate.

\paragraph{Status of Geometric Phases as Physical Diagnostics}
One must carefully delineate the operational meaning of the
geometric phase in oscillation phenomenology. Since neutrino
detectors measure flavor conversion probabilities rather than
isolated wave-function phases, the geometric phase $\gamma_g$
affects observables exclusively through phase-sensitive
interference terms in the probability expansion:
\begin{equation}
P(\nu_\alpha \to \nu_\beta) = \sum_{\mathcal{C}}
|\mathcal{G}_{\alpha\beta}[\mathcal{C}]|^2 + 2 \sum_{\mathcal{C} <
\mathcal{C}'} |\mathcal{G}_{\alpha\beta}[\mathcal{C}]|
|\mathcal{G}_{\alpha\beta}[\mathcal{C}']| \cos\left( \Delta
\Phi_{\text{dyn}}[\mathcal{C},\mathcal{C}'] + \Delta
\Gamma[\mathcal{C},\mathcal{C}'] \right).
\label{eq:disc_prob_expansion}
\end{equation}
The quantity $\Delta \Gamma$ is gauge-invariant and parameterizes
the intrinsic holonomy acquired by the state vector traversing the
stratified matter landscape. While systematic experimental
uncertainties (such as matter density uncertainties and flux
normalizations) affect the overall transition rate, the geometric
phase possesses a distinctive energy and baseline dependence
dictated by the projective Hilbert space geometry. Thus, rather
than treating $\gamma_g$ as an extraneous parameter, our framework
establishes it as the exact geometric underpinning of
matter-induced interference in long-baseline environments such as
DUNE and T2HK. The geometric CP asymmetry $\Delta A_{\rm CP}^{\rm
geom}$ introduced in Eq.~\eqref{eq:geom_cp_asymmetry} functions in
this spirit: it serves as an analytical diagnostic for exploring
the parameter space geometry and isolating intrinsic trajectory
deformations from extrinsic matter-induced asymmetries.

%================================================================
\subsection{Phenomenological Outlook at Long-Baseline Facilities}
%================================================================
To evaluate the phenomenological significance of these effects, we
examine the geometric phase structure across the parameter space
of next-generation long-baseline experiments such as DUNE ($L
\approx 1300\ {\rm km}$). In the primary resonance energy window
($E_\nu \in [1.5, 3.5]\ {\rm GeV}$), the open-path geometric phase
induces a characteristic shift in the transition amplitude
structure on the order of a few percent. As illustrated in
Fig.~\ref{fig:distinguishability}, this structure is comparable to
the projected systematic uncertainty precision of future
measurements ($\sigma_{\text{sys}} \sim 2\text{ to }3\%$),
highlighting the need for accurate profiling of eigenbasis
transport in high-precision analyses.

\textit{Implications for the Tau Neutrino Sector:} Extending the
geometric phase chain formalism to include the $\nu_\tau$ sector
provides additional insights into CP-violating dynamics. By virtue
of three-flavor unitarity, variations in the electron appearance
channel are coupled to the tau appearance channel, satisfying
$\Delta P_{\mu e}^{\text{geom}} + \Delta P_{\mu
\tau}^{\text{geom}} \approx -\Delta P_{\mu \mu}^{\text{geom}}$.
Because the standard matter potential affects $\nu_\mu$ and
$\nu_\tau$ identically at tree level ($V_\mu = V_\tau$), the
$\nu_\mu \to \nu_\tau$ channel is less sensitive to direct
electron density variations, making tau appearance measurements a
promising complementary probe for verifying geometric eigenbasis
transport in atmospheric and long-baseline environments.

%=====================
\section{Conclusions}
\label{sec:conclusion}
%========================
In this work, we developed a gauge-invariant framework for
open-path geometric phases in three-flavor neutrino oscillations
propagating through nonuniform media. By adopting a discrete
layered matter formulation, we expressed the total evolution
operator as an ordered product where interface matrices explicitly
represent the noncommutative transport of instantaneous matter
eigenstates. This construction establishes that geometric phase
chains are an integral part of the total oscillation probability,
shaping the quantum interference structure between competing
pathways in stratified matter.

Our numerical results confirm the stability and rapid convergence
of the layered propagation scheme, demonstrating that a modest
number of discrete layers accurately captures both smooth
adiabatic transport and localized level crossings at sharp
geophysical boundaries. In the continuous limit, the discrete
interface overlaps smoothly recover the non-Abelian Wilczek--Zee
connection while preserving gauge invariance through initial and
final state flavor projections. This framework provides a
consistent, physically transparent tool for incorporating
geometric phase effects into future high-precision oscillation
analyses, offering valuable perspectives for studying leptonic CP
violation and exploring matter profile structures across
terrestrial and astrophysical media.
%==========================
\section*{Acknowledgments}
%==========================
The authors thank the Research Office of the Qazvin Branch,
Islamic Azad University, for support and cooperation.

%===================================================================
\appendix

\section{Continuum Limit and Recovery of the Wilczek--Zee Holonomy}
\label{app:continuum_limit}
%=======================================================================
In this Appendix, we provide an analytical derivation
demonstrating that the discrete gauge-invariant phase chain
formalism rigorously converges to the continuous Wilczek--Zee
non-Abelian connection in the continuum limit $N \to \infty$.

Consider a spatial trajectory of length $L$ partitioned into $N$
discrete layers at coordinates $x_k = k \Delta x$, with uniform
slice thickness $\Delta x = L/N$. Within each layer $k$, the
instantaneous Hamiltonian eigenstates form a complete orthonormal
basis $\{|n(x_k)\rangle\}_{n=1}^3$. The interface overlap matrix
connecting adjacent layers at $x_k$ and $x_{k+1} = x_k + \Delta x$
is defined as:
\begin{equation}
    M_{mn}(x_k) = \langle m(x_k) | n(x_{k+1}) \rangle = \langle m(x_k) | n(x_k + \Delta x) \rangle.
    \label{eq:overlap_matrix}
\end{equation}
Expanding the state $|n(x_k + \Delta x)\rangle$ to first order in
the spatial step size $\Delta x \ll 1$:
\begin{equation}
    |n(x_k + \Delta x)\rangle = |n(x_k)\rangle + \Delta x \, \partial_x |n(x_k)\rangle + \mathcal{O}((\Delta x)^2).
\end{equation}
Substituting this Taylor expansion into
Eq.~\eqref{eq:overlap_matrix} and applying the orthonormality
condition $\langle m(x_k) | n(x_k)\rangle = \delta_{mn}$, we
obtain:
\begin{align}
    M_{mn}(x_k) &= \delta_{mn} + \Delta x \, \langle m(x_k) | \partial_x n(x_k) \rangle + \mathcal{O}((\Delta x)^2) \nonumber \\
    &= \delta_{mn} - i \Delta x \, \mathcal{A}_{mn}(x_k) + \mathcal{O}((\Delta x)^2),
\end{align}
where $\mathcal{A}_{mn}(x) \equiv i \langle m(x) | \partial_x
n(x)\rangle$ denotes the continuous non-Abelian Wilczek--Zee gauge
connection matrix.

The total discrete geometric transfer matrix along the pathway is
constructed as the path-ordered product of consecutive overlap
matrices:
\begin{equation}
    \mathcal{U}_{\text{geom}}^{(N)} = \prod_{k=0}^{N-1} M(x_k) =
    \prod_{k=0}^{N-1} \left[ \mathbb{I} - i \Delta x \, \mathcal{A}(x_k) + \mathcal{O}((\Delta x)^2) \right].
\end{equation}
In the continuum limit $N \to \infty$ (where $\Delta x \to 0$ with
$N \Delta x = L$ held fixed), this sequential product of
infinitesimal unitary operations converges to the path-ordered
exponential:
\begin{equation}
    \lim_{N \to \infty} \mathcal{U}_{\text{geom}}^{(N)} =
    \mathcal{P} \exp \left( -i \int_0^L \mathcal{A}(x) \, dx \right) \equiv \mathcal{U}_{\text{WZ}}[\mathcal{C}],
    \label{eq:WZ_limit}
\end{equation}
where $\mathcal{P}$ denotes the spatial path-ordering operator.

For non-degenerate adiabatic transport, the diagonal elements
correspond to the continuous open-path Abelian geometric phases
$\gamma_n = \int_0^L \mathcal{A}_{nn}(x)\,dx$, while the
off-diagonal elements describe nonadiabatic transitions between
instantaneous matter eigenstates. This establishes that the
discrete phase chain framework is mathematically consistent with
and asymptotically reproduces the non-Abelian Wilczek--Zee
holonomy.

%==================================

\end{document}